\documentclass[12pt,preprint]{aastex}
\usepackage{url}

\newcommand{\sat}[1]{\it\uppercase{#1}\rm}
\newcommand{\fig}[1]{Figure~\ref{#1}}
\newcommand{\tbl}[1]{Table~\ref{#1}}
\newcommand{\speed}[1]{#1 km~s${}^{-1}$}

\newcommand{\rsun}[1]{${#1}\,R_\odot$}

\begin{document}

\shorttitle{Sympathetic Filament Eruptions In One Solar Breakout Event} %

\shortauthors{Shen et al.}

\title{SYMPATHETIC PARTIAL AND FULL FILAMENT ERUPTIONS OBSERVED IN ONE SOLAR BREAKOUT EVENT}

\author{Yuandeng Shen\altaffilmark{1,2,3}, Yu Liu\altaffilmark{1,3}, Jiangtao Su\altaffilmark{3}}

\altaffiltext{1}{Yunnan Astronomical Observatory, Chinese Academy of Sciences, Kunming 650011, China; ydshen@ynao.ac.cn}
\altaffiltext{2}{Graduate University of Chinese Academy of Sciences, Beijing 100049, China}
\altaffiltext{3}{Key Laboratory of Solar Activity, National Astronomical Observatories, Chinese Academy of Science, Beijing 100012, China}

\begin{abstract}
We report two sympathetic solar eruptions, including a partial and a full flux rope eruption in a quadrupolar magnetic region, where a large and a small filament resided above the middle and the east neutral lines respectively. The large filament first rose slowly at a speed of \speed{8} for 23 minutes and then it was accelerated to \speed{102}. Finally, this filament erupted successfully and caused a coronal mass ejection. During the slow rising phase, various evidence for breakout-like external reconnection has been identified at high and low temperature lines. The eruption of the small filament started around the end of the large filament's slow rising. This filament erupted partially and no associating coronal mass ejection could be detected. Based on a potential field extrapolation, we find that the topology of the three-dimensional coronal field above the source region is composed of three low-lying lobes and a large overlying flux system, and a null point located between the middle lobe and the overlying antiparallel flux system. We propose a possible mechanism within the framework of the magnetic breakout model to interpret the sympathetic filament eruptions, in which the magnetic implosion mechanism is thought to be a possible linkage between the sympathetic eruptions, and the external reconnection at the null point transfers field lines from the middle lobe to the lateral lobes and thereby leads to the full (partial) eruption of the observed large (small) filament. Other possible mechanisms are also discussed briefly. We conclude that the structural properties of coronal fields are important for producing sympathetic eruptions.
\end{abstract}


\section{INTRODUCTION}
The filament eruption is one of the most spectacular phenomena on the Sun. Previous studies have shown that filament eruptions are closely associated with magnetic flux cancellation and emergence on the photosphere \citep{yan11}, flares, and coronal mass ejections (CMEs), which are large-scale expulsions of mass and magnetic field from the Sun into the heliosphere \citep[for a review, see][]{huds01}. It has been widely accepted that the filament eruption, flare, and CME are three different manifestations of a single physical process originating from the coronal magnetic field, but the relationships among them are not well understood \citep{lin03}. Generally speaking, the eruption of a filament always starts from a closed magnetic system in quasi-static equilibrium, in which the upward magnetic pressure force of the low-lying sheared field is balanced by the downward tension force of the overlying field. When the eruption begins, the equilibrium is destroyed catastrophically, and part of the nonpotential magnetic flux and the plasma contained within it are expelled from the Sun violently. However, when and how the eruption takes place is still not well understood, even though extensive observational and theoretical studies have been made in the past several decades. From the space weather point of view, it is important to enrich our knowledge about the physical mechanisms of various kinds of eruptive solar phenomena to improve our ability to forecast space weather. For the present, investigating the coronal magnetic field configuration and its spatial and temporal evolution with high temporal spatial resolution data are important to provide diagnostics on various stages of solar eruptions.

Previous studies have shown that filament eruptions exhibit various eruptive behaviors, including failed eruption \citep[e.g.,][]{ji03,alex06,liu09,shen11,mroz11,kuma11}, partial eruption \citep[e.g.,][]{gilb00,gilb01,zhou06,liur07,trip09}, and full eruption \citep[e.g.,][]{plun00}. \cite{gilb07} have made a specific definition about the three types of the filament eruptions. According to their definition, a failed filament eruption is a case in which neither the lifted filament mass nor the supporting magnetic structure escapes the solar gravitational field. Recent observational studies indicated that the failed filament eruptions could possibly be sub-divided into two subtypes: One is that the filament, often strongly kinked, erupts symmetrically below a group of overlying loops. When the erupting filament reaches a maximum height, it will fall back to the solar surface \citep[e.g.,][]{ji03,mroz11}. The other is that the filament erupts asymmetrically with respect to the overlying confining loops near either footpoint of the loops. When the erupting filament reaches up to the apex of the overlying loops, it stops there for a relatively long time and then falls down along one or both legs of the confining loops \citep[e.g.,][]{liu09,shen11}. Up to the present, several important impact factors are proposed to be the reasons for failed filament eruptions. For instance, a slowly decreasing gradient of the overlying magnetic field \citep{toro05}, strong magnetic field intensity at low altitude \citep{liu08}, an asymmetric confinement of the background field above the filament \citep{liu09}, and lower energy released than that of a successful eruption in the low corona \citep{shen11}.

A partial filament eruption occurs when a reconnection site is located within or above the filament. The reconnection separates the flux rope structure into two parts: an escaping and a surviving part. The former often can be identified as a CME with or without a bright core in white-light coronagraph observations, resting on where the reconnection site is. On the other hand, the latter often leads to the reformation of a new filament in the same channel shortly after the eruption. In a statistical study presented by \cite{gilb00}, they found that the separation in partial filament eruptions often occurs in the height range from 1.20 to \rsun{1.35}. \cite{gilb01} proposed a two-dimensional scenario to interpret partial filament eruptions. They suggested that the reconnection site at different positions of the flux rope would lead to different types of filament eruptions. In this scenario, if the reconnection site is located in (above) the filament, it will lead to a CME with (without) a three-part structure; if the reconnection site lies below the filament, it will lead to a successful one. A corresponding three-dimensional simulation of the partial expulsion of a flux rope was performed by \cite{gibs06}. A full filament eruption is a case in which the bulk of the filament material and the associated magnetic structure fully escape from the Sun. Such eruption often causes a three-part structure CME observed in white-light, which comprises a bright leading shell of material surrounding a dark cavity within which bright prominence material is found \citep{crif83}.

A number of theories have been developed to account for filament/CME eruptions \citep[for reviews, see][]{forb00,lin03,forb06,miki06,forb10}. Among many filament/CME models proposed so far, a majority of studies only considered the local bipolar region and did not take into account any possible multipolar component of the field on larger scales. However, many filament eruptions often occur in multipolar topologies, such as quadrupolar and $\delta$ sunspot regions \citep[e.g.,][]{liu03,ster04b}. A numerical model, called magnetic breakout model, involving multipolar field geometry was proposed to interpret filament eruptions and CMEs \citep{anti98,anti99}. This model assumes a large-scale quadrupolar field configuration, in which the core field is assumed to be increasingly sheared by photospheric motions. The core field is also surrounded by an overlying antiparallel loop system that restrains the core from erupting. Within such a magnetic configuration, there is a coronal null point that located between the core and the overlying fields. The expansion of the core will lead to a current sheet formed around the null point region, and the subsequent reconnection in this current sheet progressively removes the field lines that confine the core field by transferring them to the lateral loop systems and thereby allows the sheared core field to erupt explosively outward. In recent years, the magnetic breakout model has been enriched to a large extent \citep{lync04,lync08,macn04,devo05,devo08,phil05,zhan05,zucc08,van09}, and many observational studies have also been put forward in support of the breakout model \citep{aula00,ster01b,ster01a,ster04a,ster04b,wang02,mano03,maia03,gary04,will05,bong06}.

When multiple solar eruptions occur consecutively within a relatively short time period in one complex \citep{liuc09}, or different active regions over a large distance \citep{zhuk07,jian11}, we call them ``sympathetic eruptions''. A critical issue for sympathetic eruptions is whether the close temporal correlation between the consecutive eruptions is purely coincidental, or causally linked. The key to resolve this issue is to find the physical relationship between them. Previous statistical and detailed case studies indicated that physical connections between them do exist \citep[e.g.,][]{moon02,moon03,wang01,wang07,jian08,jian11,yang12}. At present, it seems most likely that the basic mechanism that connects sympathetic eruptions is of a magnetic nature. Very recently, \cite{schr11} investigated a series of flares, filament eruptions, CMEs, and related events that occurred on 2010 August 1--2. They found that all events during the two days were connected by a system of separatrices, separators, and quasi-separatrix layers, which indicates the importance of the magnetic topological structure of the global coronal field in the production of sympathetic eruptions. \cite{toro11} also investigated three consecutive filament eruptions that occurred on 2010 August 1. In this chain of events, one filament was located adjacent to a pseudo-streamer that contained two other filaments. The filament outside the streamer erupted first, and then the other two filaments also erupted consecutively. By performing a three-dimensional simulation of a configuration similar to the observed event, they successfully reproduced the consecutive filament eruptions, and found that the filament eruptions within the streamer resulted from the reduction of the stabilizing flux above them due to the reconnections within the streamer, triggered by the disturbance of the nearby filament eruption.

Here, we present a partial and a full filament eruption that occurred simultaneously in two neighboring source regions that produced a large-scale quadrupolar configuration. This event occurred on 2011 May 12 and was accompanied by a {\em GOES} C1.8 flare and a CME with a three-part structure. To our knowledge, this kind of event has not yet been reported before. Combining the high temporal and spatial observations taken by {\em Solar Dynamic Observatory} ({\em SDO}) and {\em Solar Terrestrial Relations Observatory} \citep[{\em STEREO};][]{kais08}, we can investigate the sympathetic filament eruptions in great detail. Based on our analysis results, we propose a possible mechanism to interpret the sympathetic filament eruptions studied in this paper, and other possible mechanisms are also discussed briefly. In the rest of the paper, instruments and data sets used in the present study are introduced in Section 2, and results are presented in Section 3. In Section 4, we propose a scenario to interpret our observational results, and discussions are also presented. Summary and conclusions are presented in the last section.

\section{INSTRUMENTS AND DATA SETS}
The Atmospheric Imaging Assembly \citep[AIA;][]{leme11} onboard {\em SDO} has high time cadences up to 12 s and short exposures of 0.1--2 s. It captures images of the Sun's atmosphere out to \rsun{1.3} with a pixel width of $0\arcsec.6$ in seven EUV and three UV visible wavelength bands. The Helioseismic and Magnetic Imager \citep[HMI;][]{scho10} onboard {\em SDO} provides line-of-sight magnetograms at 45 s cadence with a precision of 10 G.

The Extreme Ultraviolet Imager \citep[EUVI;][]{wuel04} of the Sun Earth Connection Coronal and Heliospheric Investigation \citep[SECCHI;][]{howa08} onboard {\em STEREO} provides full-disk 171 \AA, 195 \AA, 284 \AA, and 304 \AA\ images continuously with a pixel width of $1\arcsec.6$, and it observes the Sun out to \rsun{1.7}. The 171 \AA\ and 284 \AA\ images are taken with a time cadence of 2 hours, and the time cadence for 195 (304) \AA\ is 3 (10) minutes. The COR1 instrument of SECCHI onboard {\em STEREO} is an internally occulted coronagraph. It observes the inner solar corona in white-light from 1.4 to \rsun{4} \citep{thom03}. The COR1 images are taken with a time cadence of 5 minutes, and have a pixel size of $7\arcsec.5$. On 2011 May 12, the event was observed by {\em STEREO} Behind ({\em STEREO}-B), and the separation angle of {\em STEREO}-B with {\em SDO} was about $94^{\circ}$. The source region of the event presented here was near the northwest limb on the {\em STEREO}-B images, while it was near the northeast limb from the {\em SDO} viewpoint. For convenience, we will refer to the instruments {\em STEREO}-B/EUVI and COR1 as EUVI-B and COR1-B, respectively.

The H$\alpha$ data are taken by Kanzelh$\rm \ddot{o}$he Solar Observatory (KSO) in Austria, which provides H$\alpha$ full-disk images with a 1 minute cadence and a pixel width of roughly $1\arcsec$. The {\em Reuven Ramaty High Energy Solar Spectroscopic Imager} \citep[{\em RHESSI};][]{lin02} hard X-ray (HXR) sources (6--12 keV) are reconstructed using the Pixon algorithm \citep{metc96}, which provides a significantly better photometry measurement, accurate position estimation, and allows for the detection of fainter sources. The detectors 3--8 are used in our reconstruction and the integration time is 1 minute around the {\em RHESSI} flare peak. In addition, the {\em Geostationary Operational Environmental Satellite} (\sat{GOES}) soft X-ray (SXR) flux is also used.

\section{RESULTS}
\begin{figure}\epsscale{0.9}
\plotone{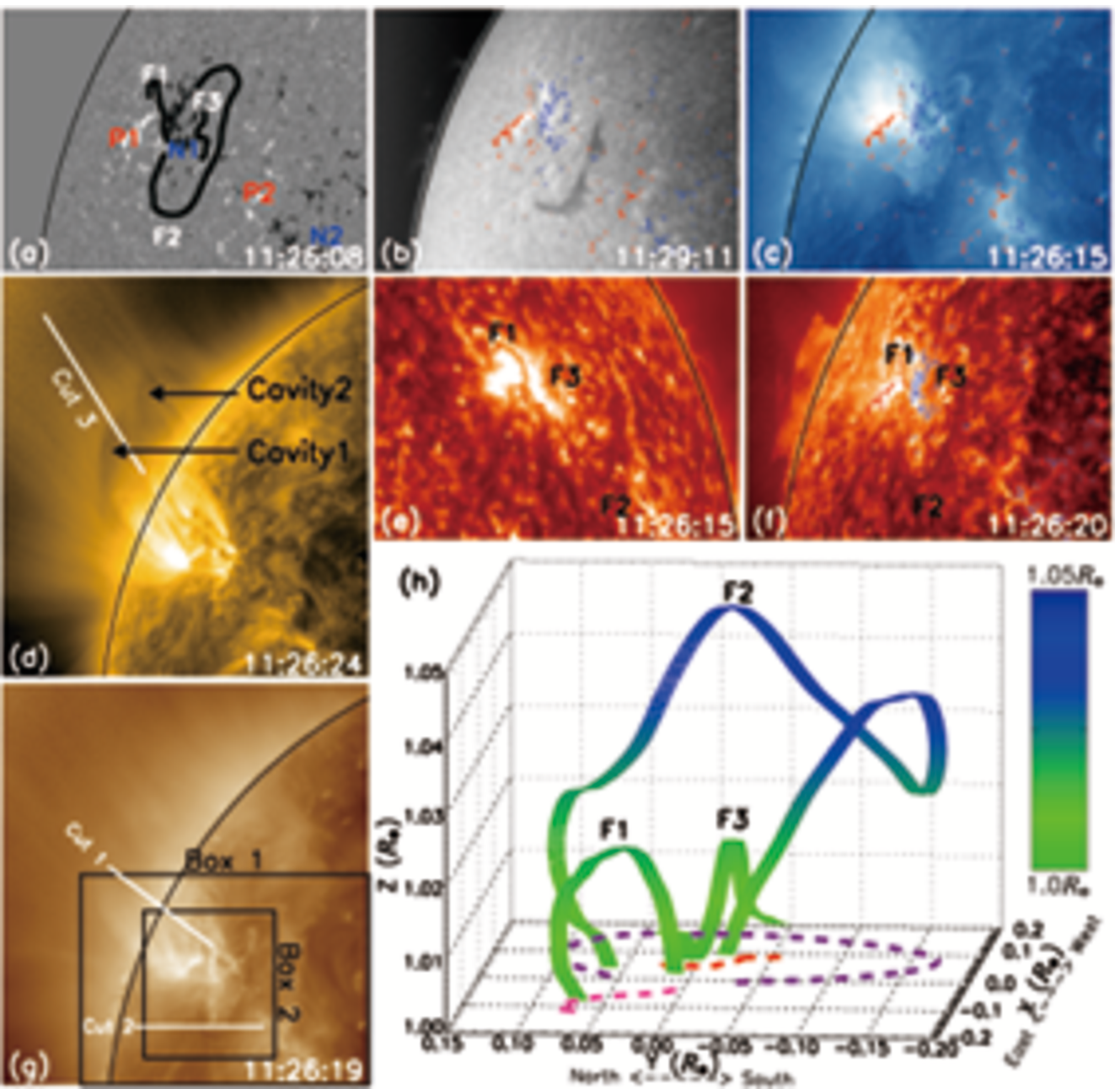}
\caption{(a) HMI line-of-sight magnetogram, (b) KSO H$\alpha$ center, (c) AIA 335 \AA, (d) AIA 171 \AA, (e) EUVI 304 \AA, (f) AIA 304 \AA\ and (g) AIA 193 \AA\ images show the event before the eruption at various wavelength bands. The filament spines detected from the AIA 304 \AA\ image at 11:26:20 UT are overlaid on the HMI line-of-sight magnetogram, and labeled as ``F1'', ``F2'', and ``F3''. The contours of the HMI line-of-sight magnetogram at 11:26:08 UT are overlaid on panels (b), (c) and (f). The contours levels are $\pm100$, $\pm300$, $\pm500$, and $\pm700$ G, with red (blue) color for positive (negative) polarity. Box 1 (2) in panel (g) indicates the FOV of panels (a)--(c), and (f) (of \fig{fig2}). The 3D-box in panel (h) shows the three-dimensional shapes of the filaments reconstructed from 304 \AA\ images shown in panels (e) and (f). The color depth represents the height of the filaments, and the pink, purple, and red dashed lines show the projection of ``F1'', ``F2'', and ``F3'' on the X--Y plane respectively. The black curve in each panel indicates the solar disk limb (the same in the subsequent figures). The field of view (FOV) for panels (a)--(c) and (f) is $550\arcsec \times 400\arcsec$, it is $700\arcsec \times 765\arcsec$ for panels (d) and (g), and $680\arcsec \times 410\arcsec$ for panel (e). \label{fig1}}
\end{figure}

The event of sympathetic eruptions on 2011 May 12 included a partial and a full filament eruption, a {\em GOES} C1.8 flare, and a CME observed by COR1-B. The primary magnetic configuration is a large-scale quadrupolar magnetic region, which evolved from the active regions 11193 and 11191 that were on the disk center on 2011 April 18. An overview of the configuration before the eruptions is displayed in \fig{fig1}. Panel (a) shows the magnetic configuration of an HMI line-of-sight magnetogram. The positive (negative) polarities are labeled ``P1'' and ``P2'' (``N1'' and ``N2''), respectively. By overlaying the filament spines on the magnetogram, one can see that F1 (F2 and F3) was located on the neutral line between P1 and N1 (N1 and P2). Two loop systems, one connecting P1 and N1 and the other connecting P2 and N2, are obvious in AIA EUV wavelengths (see panels (c), (d), and (g)). In addition, there are two cavities that can be identified off the disk limb (panel (d)).

With the procedure ``scc\_measure.pro'' developed by W. Thompson, we reconstruct the three-dimensional shapes of the spines of the three filaments by using a pair of 304 \AA\ images from EUVI-B at 11:26:15 UT and AIA at 11:26:20 UT. Obviously, F2, exhibiting a snaky structure, is located higher than F1. From the three-dimensional reconstruction, we obtain that the height of the highest section of F1 (F2) is about 16 (31) Mm above the photosphere. According to a statistical study presented by \cite{liuk12}, the height of F2 is close to the lower threshold of the critical unstable height (41 Mm) of disrupted filaments, and F1 should be more stable than F2. Since F3 and cavity 2 remained stable during the eruptions of F1 and F2, and therefore they may not belong to the magnetic system involved in the sympathetic eruptions studied here, we only focus on the evolution of F1, F2, and cavity 1. In addition, a time line of the main sub-events and the corresponding features involved in the present event can be found in Table \tbl{tb}.

\subsection{THE PRE-ERUPTIVE PHASE}
\begin{figure}\epsscale{0.9}
\plotone{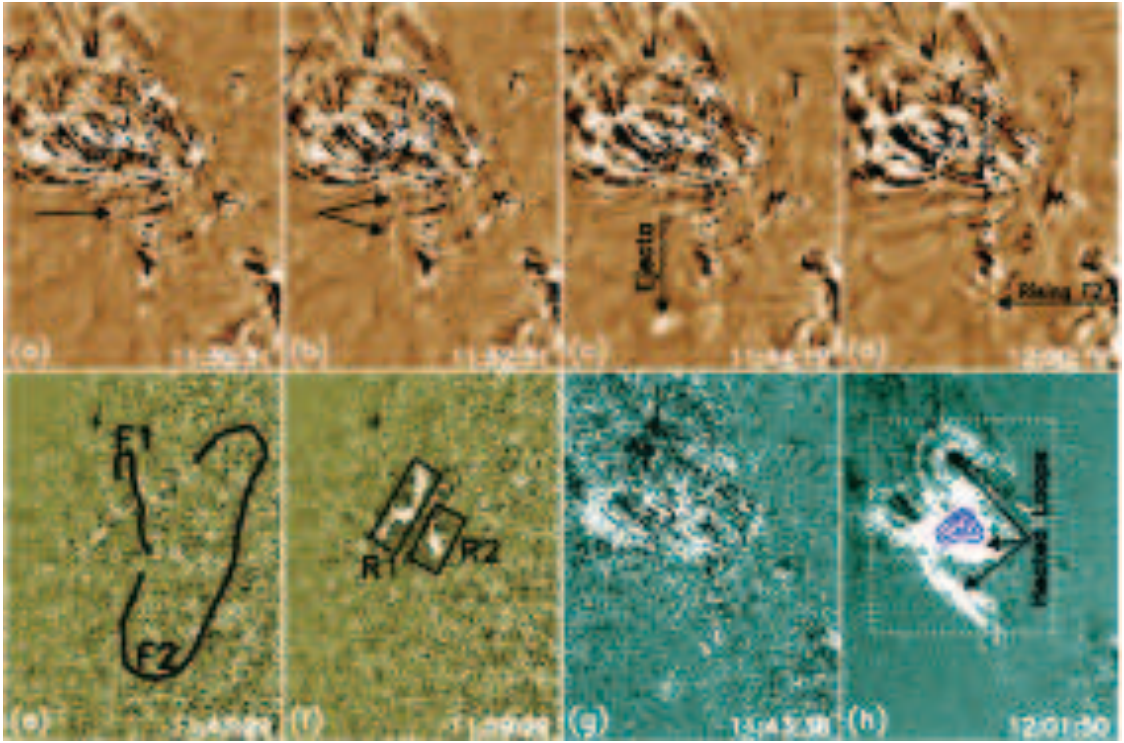}
\caption{(a)--(d) AIA 193 \AA, (e)--(f) AIA 1600 \AA, (g)--(h) AIA 94 \AA\ fixed-base difference images show the signatures observed before the main filament eruptions. The spines of F1 and F2 are overlaid on panel (e). The blue contours overlaid on panel (h) are the HXR source at 11:56 UT, and the contour levels are 30\%, 50\%, and 75\% of the maximum brightness. The arrows in panels (a)--(c) indicate the plasma ejection, while the horizontal arrow in panel (d) indicates the rising part of F2. The arrows in panel (h) point to the brightened loops on AIA 94 \AA\ images. The FOV for each panel is $250\arcsec \times 330\arcsec$. Animations 1--3 are available for this figure in the online version of the journal. \label{fig2}}
\end{figure}

Before the sympathetic filament eruptions, a small plasma ejection was observed near the southern end of F2. The start time of the ejection was at about 11:30 UT, and then it split into two parts at around 11:32 UT (see \fig{fig2}(a) and (b)). The northern part moved northward into the active region, while the southern part traveled southward and then interacted with the southernmost section of F2 at about 11:40 UT. Subsequently, F2 rose slowly from the interaction region, most likely due to the perturbation induced by the plasm ejection (see \fig{fig2}(c)--(d) and \fig{fig5}(b), and Animation 1 available in the online version of the journal). The slow rising of F2 lasted for about 23 minutes (11:45 UT--12:08 UT) at a speed of about \speed{8} in the plane of the sky. During this period, three other interesting phenomena are observed: (1) the brightening in the 1600 \AA\ images, which were located at the both sides of F1 (see \fig{fig2}(e)--(f) and Animation 2), (2) the appearance of bright loops connecting the two brightening in the AIA 94 \AA\ wavelength (see \fig{fig2}(g)--(h) and Animation 3), (3) a {\em RHESSI} HXR source detected at 11:56 UT over the bright loop top, located also between the two ribbon-like brightening (see \fig{fig2}(h)). The peculiar temporal and spatial relationship among these pre-eruption features indicates that they were produced by the same reconnection event.

\begin{figure}\epsscale{0.8}
\plotone{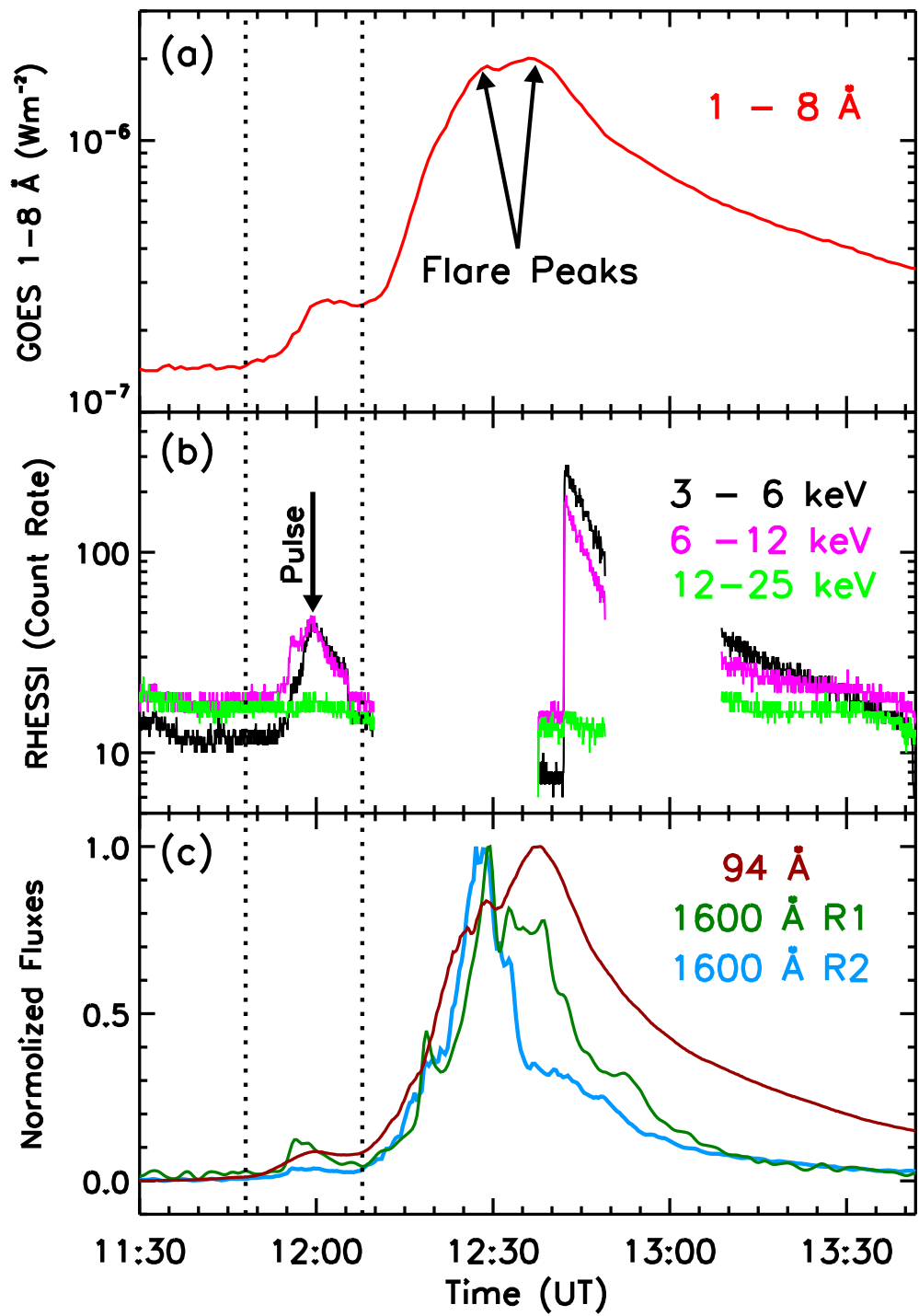}
\caption{Top panel: Time profile of {\em GOES} 1--8 \AA\ SXR; Middle panel: {\em RHESSI} HXR count rates in the energy bands (4 seconds integration) of 3-6 keV (black), 6--12 keV (magenta), and 12--25 keV (green); Bottom panel: Light curves of 94 \AA\ (darkred) and 1600 \AA (green and blue). The 94 \AA\ light curve is measured from the white dashed box region shown in \fig{fig2} (h), while the 1600 \AA\ light curves are measured from the two black box regions shown in \fig{fig2} (f). The two arrows in the top panel point to the two peaks of the {\em GOES} C1.8 flare, while the vertical arrow in the middle panel indicates the pulse just before the main flare. \label{fig3}}
\end{figure}

\fig{fig3} shows the time profiles of the {\em GOES} 1--8 \AA\ SXR flux, the {\em RHESSI} HXR count rates in the lower energy bands, and light-curves of the two brightening regions in 1600 \AA\ and the region of the hot loops in the 94 \AA\ images (see the boxes in \fig{fig2}(f) and (h)). The time profiles of {\em GOES} 1--8 \AA\ flux show a small hump (peaking at about 12:00 UT) just before the onset of the main C1.8 flare, which peaked at 12:28 UT and 12:37 UT respectively (indicated by the two arrows in \fig{fig3}(a)). The {\em RHESSI} HXR count rates and the light-curves of 1600 \AA\ and 94 \AA\ also exhibited such a hump around the same time (see \fig{fig3}(b) and (c)). It should be noted that these phenomena are detected during the slow rise stage of F2 just before the filament eruptions. The temporal and spatial relationship indicates that they are probably the manifestations of the external reconnection as predicted in the magnetic breakout model \citep[see,][for detail]{anti99}.

\subsection{THE ERUPTIVE PHASE}
\begin{figure}\epsscale{0.9}
\plotone{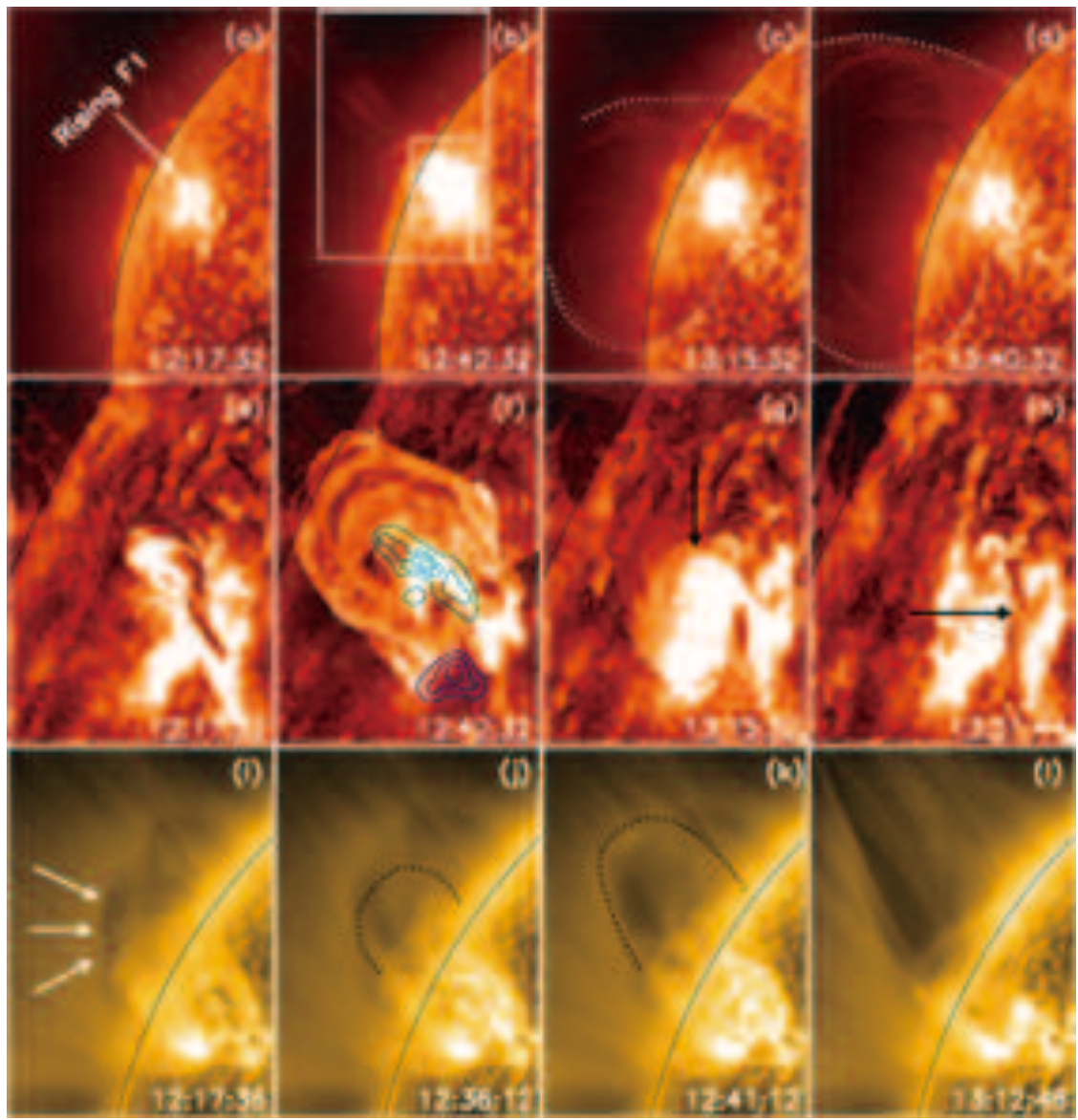} \caption{AIA 304 \AA\ ((a)--(h)) and 171 \AA\ ((i)--(l)) images show the eruptions of F1, F2, and the blob-like feature above F1. The small (large) white box shown in panel (b) indicates the FOV of the middle (bottom) row of the figure. The white dotted curves in panels (c) and (d) mark the outer profile of the erupting F2. The blue and cyan contours overlaid on panel (f) are the HXR sources at 11:56 UT (blue) and 12:43 UT (cyan), and the contour levels are the same with \fig{fig2}. The arrow in panel (g) points to the wedege-like post-eruptive loop structure, while the arrow in panel (h) points to the reformed filament. The white arrows in panel (i) indicate the outer profile of cavity 1, and the black dotted curves mark outer profile of the erupting blob-like feature. The FOV for the top, middle, bottom rows are $620\arcsec \times 860\arcsec$, $160\arcsec \times 220\arcsec$, $400\arcsec \times 580\arcsec$, respectively. Animations 4--6 are available for this figure in the online version of the journal. \label{fig4}}
\end{figure}

\begin{figure}\epsscale{0.9}
\plotone{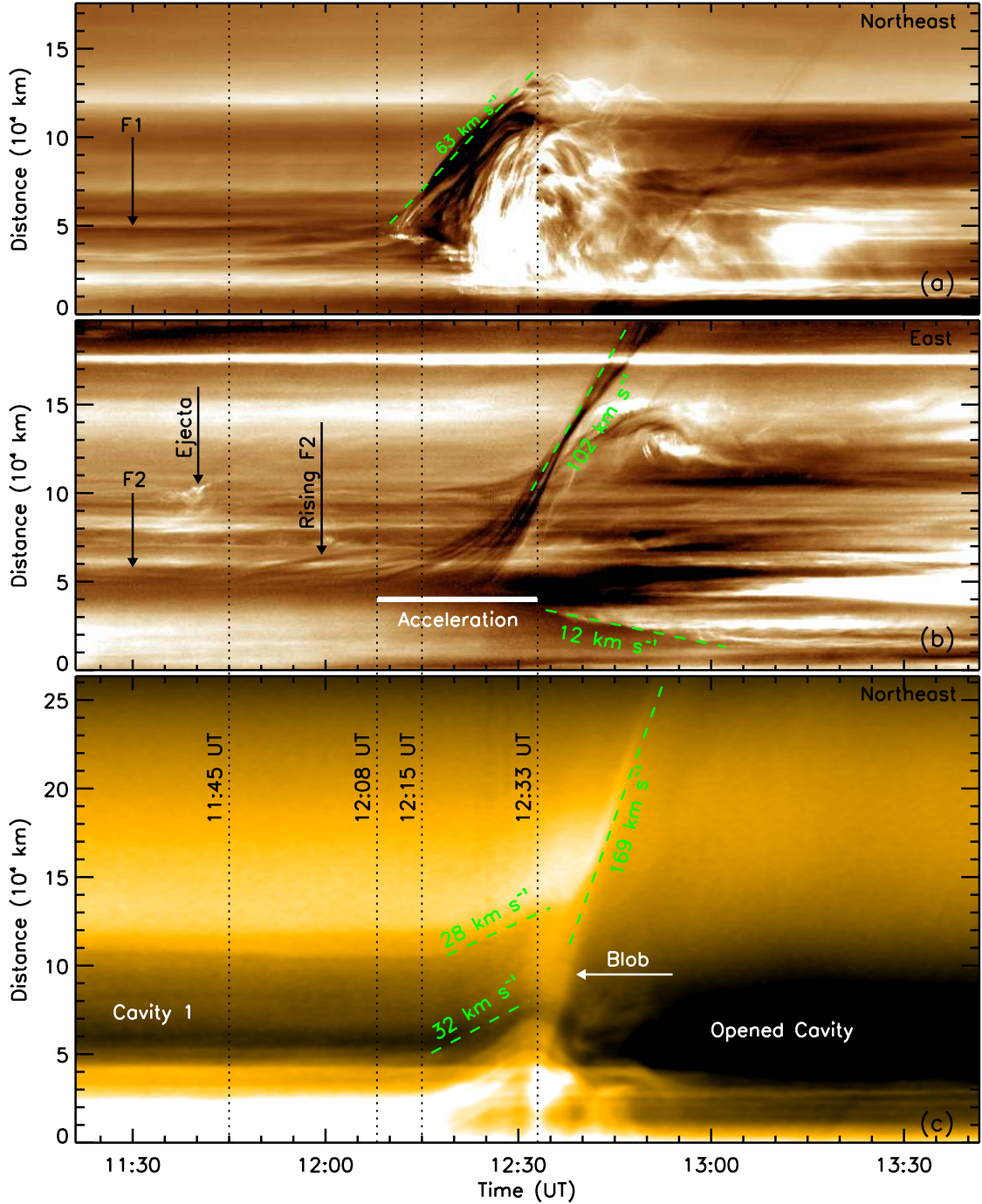} \caption{Time-distance diagrams obtained from cut 1 (for F1), cut 2 (for F2), and cut 3 (for the cavity) shown in \fig{fig1}. The vertical dotted lines indicate the four critical time points. Green dashed lines are the linear fit to the moving features and the corresponding speeds are also plotted. The horizontal white bar in panel (b) marks the acceleration stage of F2, and the white arrow in panel (c) indicates the blob erupting. It should be noted that the white horizontal stripe in the top of panel (b) is a bright point rather than the disk limb. \label{fig5}}
\end{figure}

The simultaneous eruptions of F1, F2, and of a blob-like structure above the apex of the erupting F1 are all displayed in \fig{fig4}, and their detailed kinematics are studied with time-distance diagrams as shown in \fig{fig5}. The top row of \fig{fig4} shows the full view of the eruptions of F1 and F2 using AIA 304 \AA\ images, while the middle row is a close-up view on the erupting F1. The eruption of F2 showed clear untwisting motions (see Animation4) and experienced three eruption phases: the slow rising phase (11:45--12:08 UT, $v=$\speed{8}), the acceleration phase (12:08--12:35 UT, $a=13.25$ m s$^{-2}$), and the fast eruption phase (after 12:35 UT, $v=$\speed{102}; see \fig{fig5}(b)). Just after the slow rising period of F2, F1 began to rise with a constant speed of about \speed{63}. The eruption of F1 showed a nice kinked-shape and exhibited obvious writhing motions during this period (see \fig{fig4}(e) and Animation 5). After F1 reached the maximum height ($\sim 90$ Mm above the solar surface) at around 12:33 UT, the erupting material started to drain back to the solar surface. Around 12:20 UT, a strong emission occurred around the intersection of the F1's two legs, and a {\em RHESSI} HXR source is also found at this site (the {\em RHESSI} data available at 12:42 UT, see the cyan contours in \fig{fig4}(f)). This result may indicate the occurrence of reconnection in a current sheet underneath the apex of the writhing F1, which may formed via a kink instability resulting from the interaction of the two adjacent legs underneath the apex of the writhing filament \citep[e.g.,][]{fan04,alex06,liur07,klie10}, or by the instability-induced collapse of an X-line or hyperbolic flux tube that runs below the rising F1 \citep[e.g.,][]{toro04}. In the falling stage of F1, a bright wedge-like post-eruptive loop structure can be observed (see \fig{fig4}(g)). After the eruption, the filament reappeared at the same place where F1 had resided in (see \fig{fig4}(h)). The evolutionary process of F1 is reminiscent of the failed filament eruption on 2002 May 27 that has been studied in detail by \cite{ji03}, \cite{toro05}, and \cite{alex06}. In their case, the filament experienced a kinking motion that was effectively stopped at a height of $\sim 80$ Mm above the photosphere, and an HXR source was also identified under the apex of the strongly kinked filament during the erupting phase \citep{alex06}. In addition, the reformation of the filament suggests that the eruption of F1 was possibly a partial filament eruption \citep{trip09}. Moreover, as can be seen in the AIA 171 \AA\ images, an intriguing blob-like feature above the apex of the writhing F1 erupted following the eruption of F1. It first appeared at about 12:15 UT and moved along the northeast direction. Meanwhile, an obvious outward motion of cavity 1 was also observed. At about 12:35 UT, the blob-like feature interacted with cavity 1 and then they were moving outward together, leaving behind a clear dark cavity (see the bottom row of \fig{fig4} and Animation 6). The eruption of the blob-like feature and the reformation of F1 together indicate that the eruption of F1 is a partial filament eruption, in which the reconnection possibly occurs above the apex of the erupting F1 \citep[also see,][]{trip09,liur07}.

To make clear the temporal relationship among the eruptions of F1, F2, and the blob-like feature, we plot the time-distance diagrams of them in \fig{fig5}, in which the main features are indicated by arrows, the important times are highlighted by the four vertical dotted lines, and the speeds of the moving features in the plane of the sky are also plotted (see the green dashed lines that show linear fit to the stripes). Except for the features described in above paragraph, it is interesting that an obvious moving flare ribbon with a speed of \speed{12} was identified at the beginning of the fast eruption phase of F2 (see \fig{fig5}(b)), which may indicate the rising of the reconnection site that is associated with the erupting F2. The conjugate ribbon can also be found near the active region, and the moving speed is obviously smaller than the western ribbon (see Animation 5). In addition, obvious post-eruptive flare loops could be identified on the AIA and EUVI-B images during this phase. These signatures suggest the reconnection process associated with the erupting F2, which is similar to the physical picture described in the catastrophic flare/CME model by \cite{lin00}. It is worth noting that the start time of the activation of F1 was at around 12:03 UT, which was close to the phase transition time of F2 from the slow rising phase to the acceleration phase.

This close temporal relationship suggests a close relation between F1 and F2. Since the rising of F2 occurred before the activation of F1, it seems that the eruption of F2 reconfigured its surrounding magnetic condition and thereby triggered its neighboring F1. In addition, as mentioned above, the {\em GOES} C1.8 flare showed double peaks at 12:28 and 12:37 UT. Considering the time of the flare and the eruptions of F1 and F2, we infer that the flare possibly resulted from two components: one was related to the reconnection underneath the erupting F1, and the other to the moving ribbon, related to the reconnection below the erupting F2. However, the moving ribbon associated with F2 was very weak compared to the emission caused by F1. Therefore, there is another possibility that both the double peaks of the flare were caused by the strongly heated filament material of F1 during its eruption, in which the emission will come from both the heated filament material and from the related chromospheric ribbons, and their contributions may not be fully simultaneous and therefore produce the double flare peaks observed by {\em GOES}.

Along with the rising of F1, a blob-like loop structure intruded into cavity 1 at about 12:15 UT with a speed of about \speed{32}. It was also accompanied by the expansion of cavity 1 at a speed of about \speed{28}, slightly smaller than that of the blob. At about 12:33 UT, when F1 reached its maximum height, the majority of the risen loops that confined F1 started to fall back, but the blob kept on moving outward and subsequently interacted with cavity 1, and then they erupted together with a speed of about \speed{169} (see the bright convex structure in the lower part of \fig{fig5}(c)). For the blob eruption, its transition from the slow rising phase to the fast eruption phase occurred when F1 reached its maximum height, which suggests that the blob was separated from the flux rope structure that contained F1 due to reconnection that occurred within the flux rope system. Since no filament material expulsion could be detected in the eruption of F1, we conjecture that the reconnection site was located about the apex of F1.

\begin{figure}\epsscale{0.9}
\plotone{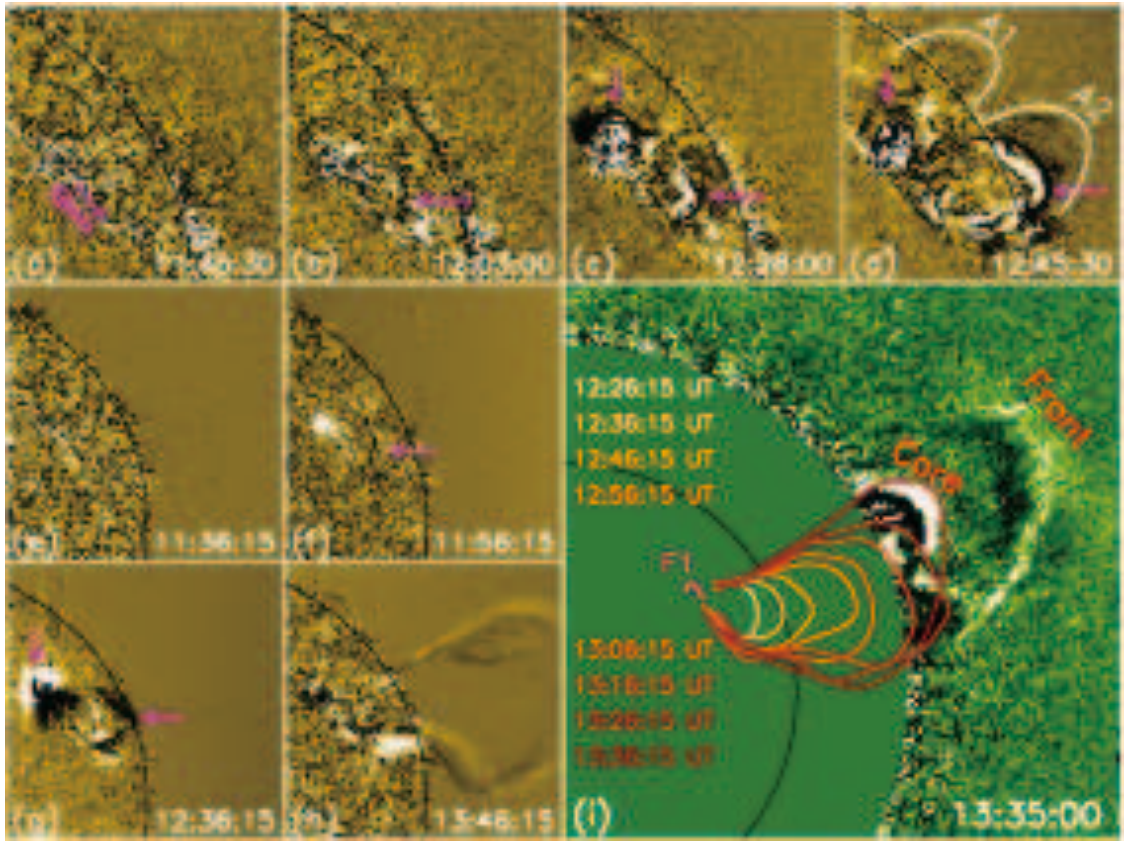} \caption{EUVI-B 195 \AA\ ((a)--(d)), 304 \AA\ ((e)--(h)), and COR1 (i) running difference images show the eruption of F1, F2, and the associated CME. The vertical pink arrows point to F1, while the horizontal pink arrows point to F2. In addition, the pink arrow in panel (a) points to the ejection. The two white arches in panel (d) mark the preceding arches above the erupting F1 and F2. The spines of the erupting F2 are overlaid on the COR1 image (panel (i)), with different colors representing different times. The FOVs for panels (a)--(d) and (e)--(h) are $830\arcsec \times 830\arcsec$ and $1100\arcsec \times 1100\arcsec$, respectively.  Animations 7--9 are available for this figure in the online version of the journal. \label{fig6}}
\end{figure}

From another angle, {\em STEREO}-B also catched this event, and the results are shown in \fig{fig6}. In the EUVI-B and COR1-B observations, except for the cavities, the main features, such as the plasma ejection, the consecutive eruptions of F1 and F2, and the post-eruptive flare loop related to F2, can also be observed. (see Animations 7--9). An intriguing phenomenon is that each filament was encircled by a preceding bright arch during their eruptions. For convenience, we call the arches preceding F1 and F2 A1, and A2 respectively (see the white curves in \fig{fig6}(d)). A2 is possibly the progenitor of the bright front of the CME recorded by COR1-B (\fig{fig6}(i)). It is clear that the CME was caused by the erupting F2. By overlying the spines of the erupting F2 at different time on the COR1-B image at 13:35 UT, a close temporal and spatial relationship between the CME and the erupting F2 can be found (see \fig{fig6}(i)). Obviously, the bright core of the CME corresponds to the erupting F2, while the CME's bright front is the development of the erupting A2 on the white-light COR1-B observations. In addition, the erupting F2 showed a northward drift relative to the initial eruption direction, which probably resulted from the orientation of the cavity overlying it. However, we do not find a CME that could be considered as the developing of the erupting A1, which is possibly because A1 is too faint to be detected by COR1. By comparing the observations of {\em STEREO}-B and {\em SDO}, we infer that the erupting A1 observed at EUVI-B 195 \AA\ should be the counterpart of the erupting blob-like feature on the AIA 171 \AA\ images.

\subsection{PFSS EXTRAPOLATION AND DECAY INDEX ANALYSIS}
The event presented in this paper occurred near the eastern limb on the {\em SDO} images, and thus there was no reliable direct magnetic field measurement available. So we resort to the potential field source surface (PFSS) model \citep{scha69,schr03} to obtain information of the three-dimensional magnetic environment in the vicinity of the quadrupolar source region. In this model, the fields in the corona are assumed to be a potential field (current-free) and become radial at the source surface at \rsun{2.5}. The extrapolation is based on a synoptic magnetic map that was composed of a series of consecutive HMI line-of-sight magnetograms, within a limited area around the central meridian. Since the eruptions occurred on the eastern limb, the information of the source region on the synoptic map is mainly from that of three weeks ago and several days after, when the source region passed through the central meridian. In spite of this, we believe that the basic magnetic topology of the coronal field is still reliable, since the basic configuration of the source region on the HMI line-of-sight magnetograms did not change too much during the time interval of consideration.

\begin{figure}\epsscale{0.9}
\plotone{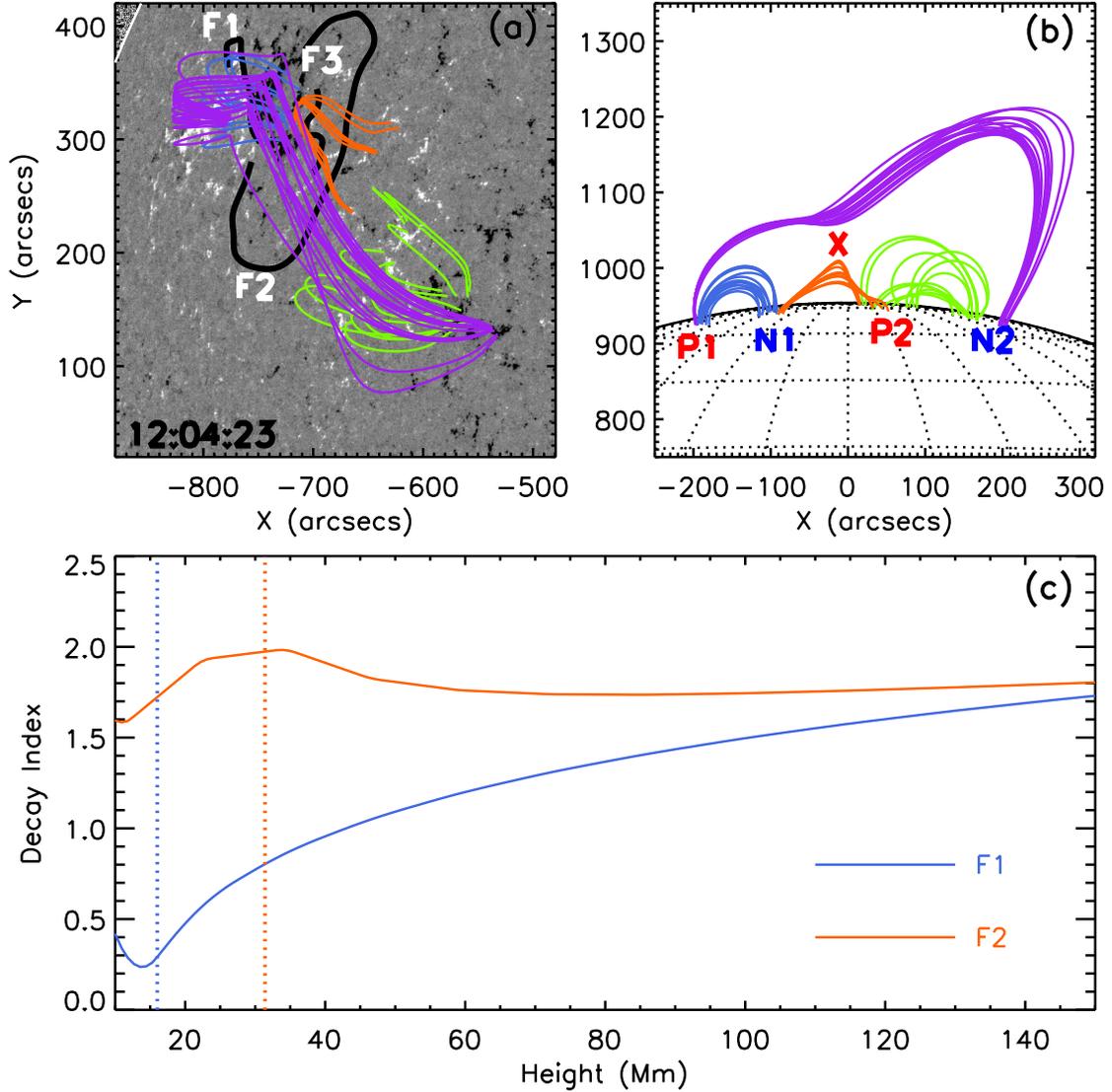} \caption{(a) HMI light-of-sight magnetogram overlaid by the magnetic field lines extrapolated from PFSS model. (b) the same magnetic lines rotated to the limb. (c) the decay indexes of the transverse magnetic field above F1 (blue) and F2 (orange). The coronal null point is labeled by the red ``X'' symbol in panel (b), and the polarities are indicated by the labels ``P1'', ``N1'', ``P2'', and ``N2'', respectively. The vertical dotted lines in panel (c) indicate the initial heights of F1 (blue) and F2 (orange) respectively. \label{fig7}}
\end{figure}

The results of the extrapolated coronal field are shown in \fig{fig7}. It should be noted that only a few representative field lines are plotted in the figure. In this figure, one can see that three low-lying lobes (blue, orange, and green) connecting the four polarities are enclosed by a group of high, large field lines (magenta) connecting P1 and N2. F1 and F2(F3) are located under the eastern and the middle low-lying lobes respectively. A coronal null point is found between the low-lying middle lobe and the overlying antiparallel flux system. The position of the coronal null point is indicated by the red ``X'' symbol in \fig{fig7}(b). Such magnetic configuration is consistent with the magnetic topology in magnetic breakout models \citep[e.g.,][]{anti99}. By comparing the coronal field extrapolated from the PFSS model and the {\em SDO} and {\em STEREO}-B observational results, it seems likely that the overlying large loop system (magenta) can be considered as the counterpart of cavity 1 observed on the AIA 171 \AA\ images, while the left and the middle low-lying lobes are possibly the initial states of the erupting A1 and A2 observed on the EUVI-B images.

Based on the calculated three-dimensional coronal fields, we examine the gradient with respect to the height of the background magnetic fields, which is believed to be an important impact factor to diagnose the ultimate fate of a filament eruption \citep[e.g.,][]{toro05,klie06,fan07,isen07,liu08,aula10,olme10,toro10,bi11,shen11}. This parameter is expressed as the so-called decay index, defined as $n = - \frac{d\log(B)}{d\log(h)}$. Here, $B$ is the strength of the external field confining the erupting core and $h$ the height above the photosphere. The theoretical instability condition is $n > 1.5-2.0$ \citep{klie06}. When the decay index satisfies this condition, a successful eruption can be expected due to the torus instability \citep{toro05,klie06} or partial torus instability \citep{olme10}. It should be noted that the external field is produced by sources other than the currents in the sheared core field, and only the poloidal component of the external field was used to calculate the decay index in the theoretical work of \cite{klie06}.

In our calculation, we use the field extrapolated by the PFSS model to approximate the real external field since it is difficult to separate the current-induced field from the PFSS model field, and the transverse component of the three-dimensional coronal field is used to calculate the decay index. We believe that this approximation has little influence on the results of this study, since we just make a comparison between the relevant values for the two filaments. At each height, the transverse component field is averaged over a small pixel region above the filament spine. \fig{fig7}(c) shows the plots of the decay index varying with height for F1 (blue) and F2 (orange). In general, the decay index of the coronal field above F1 is lower than that of F2 at each height, which suggests that the transverse magnetic field in the lower corona above F1 decreases slower than that for F2. This result is in agreement with our observational results that F1 erupted partially while F2 erupted successfully, and it is also consistent with the theoretical prediction that the decay index for a confined eruption is typically smaller than that for a successful one. It is interesting that the height distribution of the decay index for F1 increases monotonically, whereas for F2, the decay index first shows a rapid increase and then decreases to a lower constant level. Such changing pattern of the decay indexes in the low corona suggests a difference distribution of the coronal field above F1 and F2, and that the coronal field above F2 is more complex than that of F1.

\section{INTERPRETATION AND DISCUSSIONS}
\begin{figure}\epsscale{0.9}
\plotone{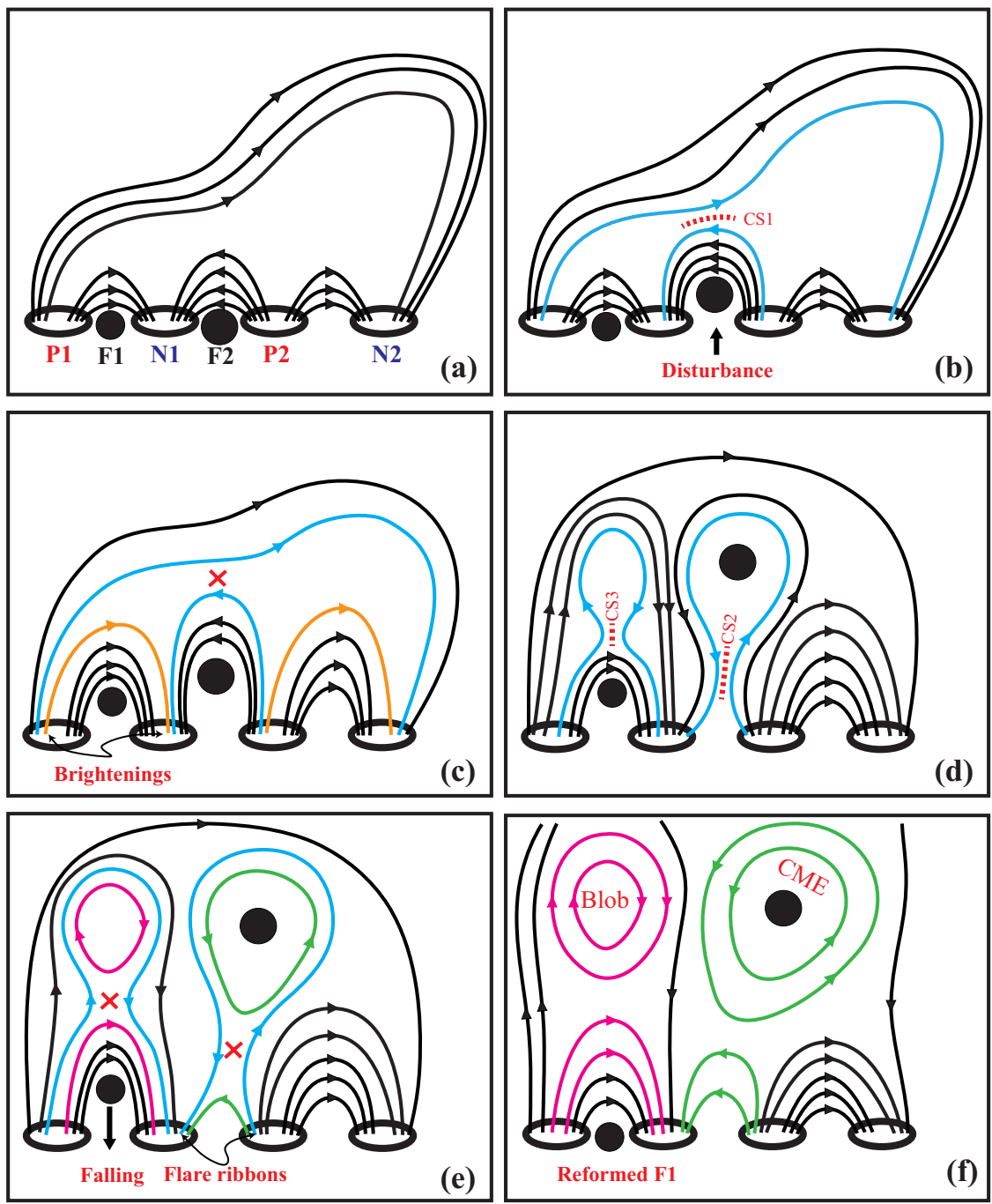} \caption{Schematic demonstrating the simultaneous filament eruptions within the quadrupolar magnetic configuration. (a) the initial magnetic configuration. (b) the rising of F2 and formation of CS1. (c) the external reconnection and the activation of F1. (d) formation of CS2 underneath F2 and an ``X'' point above F1. (e) reconnections in CS2 and CS3, and the falling of F1. (f) the reformation of F1 and the successful eruption of F2. The red dotted lines indicate location of the current sheets, while the reconnection sites are labeled by the red ``X'' symbols. The blue lines represent the field lines to be reconnected, and the orange, green, and pink lines represent the reconnected lines produced by the reconnections in the current sheets above F2, underneath F2, and above F1, respectively.  \label{fig8}}
\end{figure}

Based on our analysis results, we propose a possible interpretation for the sympathetic filament eruptions presented in this paper, within the context of the magnetic breakout model (\fig{fig8}). Panel (a) shows the initial magnetic topology of the coronal field. In this configuration, four polarities (P1, N1, P2 and N2) are connected by four groups of flux systems, and two filaments (F1 and F2) are confined by the left and middle low-lying flux systems. The initiation of the eruptions starts from the disturbance induced by a plasma ejection. Subsequently, F2 rises slowly while F1 remains stable for a certain time. The rising of F2 will lead to expansion of its overlying fluxes and formation of a current sheet (CS1) between these fluxes and their overlying large antiparallel flux system (see the red dashed curve in \fig{fig8}(b)). The continuously rising of F2 will speed up the magnetic reconnection in CS1. This reconnection will progressively remove the field lines confining F2 by transferring them to the lateral low-lying flux systems and thereby reduce the magnetic confinement ability of the fields above F2. On the other hand, the accelerated particles spiral along the reconnected field lines down to the dense chromosphere, where HXR bremsstrahlung will be created because of their interaction with the thermal ions. During this period, brightening at the footpoints of the reconnected field lines can be expected at various lines, including SXR (HXR) detected by {\em GOES} ({\em RHESSI}) instruments (\fig{fig8}(c)).

As proposed by \cite{huds00}, a magnetic implosion must occur simultaneously during the energy conversion process in coronal transients such as flares and CMEs. Under the condition of energy conservation, the energy release between the states before and after coronal transients implies the reduction of the upward magnetic pressure and thereby results in the contraction of the coronal fields overlying the energy releasing site \citep[see][for details]{huds00,liur09}. It should be noted that this magnetic implosion just occurs during the onset of the coronal transients in a short timescale (several minutes) and with a slow speed of a few \speed{}. Recently, an unambiguous case which shows large-scale contraction of the coronal loops has been reported by \cite{liur09}. In their case, the contraction sustained for about 10 minutes at a mean speed of \speed{5}, and then it was followed by the fast expansion of the coronal loops due to the compensation of energy from other regions. Other indirect evidence for the magnetic implosion phenomenon mainly comes from the observations on both converging motions of the conjugate footpoints and the descending motion of looptop sources during the early phase of some flares \citep[e.g.,][]{sui03,sui04,ji04,vero06}.

In the present case, we consider such a magnetic implosion mechanism as the linkage between the sympathetic filament eruptions, even though no obvious contraction of the overlying coronal loops could be detected from the observations. Based on the magnetic implosion conjecture, however, we believe that the reduction of the magnetic pressure around the reconnection region must occur due to the releasing of the magnetic free energy during the early stage of the external reconnection in CS1. A reduction of the magnetic pressure there will lead to the contraction of the overlying loops connecting P1 and N2, and the expansion of the three low-lying lobes. The expansion of the left lobe that connects P1 and N1 will successively reduce the stabilizing tension on F1. On the other hand, the strong writhing of F1 indicates that its eruption was driven by the kink instability within it. The reduction of the magnetic tension force will lead to the eruption of F1, by triggering the kink instability within the filament. Hence, the implosion could be a possible physical linkage of the sympathetic filament eruptions. The continuously rising of F2 will result in the formation of a new current sheet (CS2) between the legs of the field lines astride F2. The reconnection occurring in CS2 cuts the field lines that confine F2 and thus produces the observed CME, the moving flare ribbons, and the post-eruptive flare loops. The eruption of the blob and the reformation of F1 could be explained using the partial flux rope eruption scenario \citep{gilb01}, in which reconnection occurs in a current sheet (CS3) that locates above the filament. It should be noted that the reconnection in CS1 will increase the confinement ability of the fields above F1 by transferring the field lines from the middle lobe to the lateral lobes, and the reconnection in CS3 also increases the confinement of the fields above F1. Hence, the eruption of F1 is just a partial flux rope eruption and no filament material expulsion. The decay index analysis indicates that the drop of the field above F1 stayed for a considerable height range below the typical torus instability threshold, which may be an additional reason for the magnetic confinement of F1.

In our interpretation presented in \fig{fig8}, brightening should be appeared at P1, N1, P2, and N2 during the external reconnection in CS1, as the case reported by \cite{gary04}. Whereas, in the case presented here, obvious brightening only observed at P1 and N1. The absence of brightening at P2 and N2 are possibly due to the weak and dispersed distribution of the magnetic fields in these regions, as shown by the HMI magnetograms. In addition, all the brightened loops at 94 \AA\ during the slow rising phase of F2, the peculiar HXR source at 11:56 UT, and the small hump detected before the initiation of the flare on the time profiles of {\em GOES} 1--8 \AA\ SXR flux and {\em RHESSI} HXR count rates can be considered as the evidence for the external reconnection, in which the HXR source represents looptop source. Moreover, if we consider cavity 1 as the counterpart of the overlying large flux system shown in the cartoon (\fig{fig8}), an expansion motion of cavity 1 should be observed. However, the observational fact is that the obvious expansion of cavity 1 started along with the eruption of the blob, several minutes after the end of the slow rising phase of F2 (see \fig{fig5}). One possibility for this discrepancy is that the expansion of cavity 1 was too weak to be detected during the early phase of the external reconnection. The other possibility is that cavity 1 did not belong to the quadrupolar magnetic system at all, since we do not observe it on the {\em STEREO}-B observations. Both the post-eruptive flare loops and the moving flare ribbon associated with F2 could be considered to be the evidence for the reconnection in CS2. For the reconnection occurred in CS3, we do not find significant evidence for it. However, the observational results do indicate the eruption of F1 was a partial flux rope eruption, such as the reformation of F1. Further detailed studies on such partial flux rope eruptions are needed to clarify this mechanism.

The sympathetic filament eruptions presented in this paper show some similarity to the study presented by \cite{toro11}. In their case, two filaments, which are confined by a pseudo-streamer, erupt consecutively due to the removal of a sufficient amount of confining fields above them. In our case, the rising of F2 could lead to the external reconnection in CS1. This reconnection successively removes the stabilizing flux above F2, which further accelerates the eruption of the filament. It should be noted that such a mechanism is also important for producing coronal blowout jet \citep{shen12}, in which external reconnection removes the confining fluxes above a small filament close to the jet base, and thus leads to the eruption of the filament.

Both our results and the study by \cite{toro11} indicate that the structural properties of the large-scale coronal field are important for producing sympathetic eruptions. In the study performed by \cite{schr11}, they found that a series of explosive and eruptive events were connected by a system of separatrices, separators, and quasi-separatrix layers. This study highly stresses the importance of the topological features in the coronal field, and proposed three different interpretations to explain the series of consecutive eruptions \citep[see][for detail]{schr11}. One of their interpretations is that sympathetic eruptions might be the result of the destabilization of local field configurations by an overall change of the large-scale coronal magnetic field. In this scenario, the events that occur simultaneously or in a short timescale over long distances are not a chain in which one triggers another. This mechanism, to a certain extent, could be apply to explain the sympathetic filament eruptions presented in this paper. Based on our analysis results, we can assume that F1 was close to the kink instability threshold, and F2 was close to the critical height. These magnetic systems in  quasi-static equilibrium were prone to be destabilized catastrophically by the large-scale change in the global magnetic configuration. This explanation implies that the eruptions of F1 and F2 were independent, and that no direct physical linkage existed between them. However, we find various evidence that support the magnetic breakout scenario as shown in \fig{fig8}, such as the pre-eruptive signatures and the coronal null point found in the three-dimensional coronal field topology by using the PFSS extrapolation.

In addition, for the small plasma ejection observed at around 11:30 UT, it first split into two parts and then moved bidirectionally in the southern and the northern direction. While the southern part interacted with F2, the northern part traveled into the source region where F1 resided in. This raises the question whether the consecutive eruptions of F1 and F2 were triggered by the split plasma ejection simultaneously, and whether the brightening around P1 and N1 were caused by, for instance, tether-cutting-like reconnection below F1 initiated by the northward plasma ejection, and the reconnection further led to the destabilization of F1. To clear these questions, we check the temporal relationship between the plasma ejection and the pre-eruptive signatures. We find that the northward plasma ejection disappeared at about 11:38 UT, the brightening first appeared at about 11:54 UT, and the start time of the rising of F1 was about 12:08 UT. These temporal relationships indicate that the pre-eruptive signatures were possibly not triggered by the plasma ejection. On the contrary, they were possibly resulted by the external reconnection at the null point, as discussed above.

\section{SUMMARY AND CONCLUSIONS}
We present the first observational evidence for that both partial and full flux rope eruptions can simultaneously occur in one solar breakout event. The high temporal and high spatial observations of AIA allow us to distinguish the relationship between the both filament eruptions. In the event, the initiation of the large-scale magnetic system started from the perturbation induced by the plasma ejection, which directly acted on F2 and thereby led to the rising of the filament. The eruption of F2 was successful and associated with a CME observed in white-light observations, which underwent three eruption phases: the slow rising phase, the acceleration phase, and the fast eruption phase. During the slow rising phase, evidence for the external reconnection around the coronal null point are identified, such as the brightening on both sides of F1, the bright loops at AIA 94 \AA\ wavelength, the HXR source at 11:56 UT, as well as the small hump detected by {\em GOES} and {\em RHESSI} before the main C1.8 flare. On the other hand, F1 starts to erupt at around the end of the slow rising phase of F2, and the eruption of F1 was interpreted as a partial flux rope eruption, in which the flux rope was separated by the reconnection occurred above the apex of F1, and no filament material expulsion could be detected. The three-dimensional coronal field is obtained based on the PFSS extrapolation, which reveals the basic magnetic topology of coronal field above the quadrupolar source region. The magnetic configuration is composed of three low-lying lobes and a large-scale overlying flux system, in which a coronal null point is found between the middle lobe and the overlying antiparallel field lines. By calculating the decay index of the coronal fields above the two filament, we find that the decay index for F1 is smaller than that for F2 at the same height, in agreement with the theoretical prediction that the decay index for a confined eruption is typically smaller than that for a successful one. In addition, the height distribution of the decay index for F2 showed a decrease at the lower height, whereas, the decay index for F1 increases monotonically with height. The distribution of the decay indexes above the filaments in the low corona indicate the coronal field distribution above F2 is more complex than that of F1.

A possible mechanism is proposed to interpret the sympathetic filament eruptions. In our interpretation, the eruption of F2 was caused by the removal of a sufficient amount of the stabilizing field above the flux rope via the external reconnection around the coronal null point. In the meantime, the external reconnection also increases the confinement ability of the fields above F1. To link the two consecutive filament eruptions, we adopt the magnetic implosion mechanism during the early phase of external reconnection to connect them, even though we do not find obvious evidence for the implosion mechanism in our available observations. However, this interpretation is supported by a number of observational evidence, such as the various pre-eruptive signatures. We stress that the magnetic breakout scenario is just one possibility for the case presented in this paper. We do not intend to exclude other possibilities, such as the change of the large-scale magnetic configuration. In any case, the topology properties of the large-scale coronal field are important for producing sympathetic eruptions.

In summary, our analysis results support the breakout scenario proposed by S. Antiochos and coworkers, and we believe that the existence of the physical linkage between the sympathetic filament eruptions studied in this paper. More investigations involving similar magnetic structure with high temporal and spatial resolution observations would be helpful to fully understand the physical mechanisms of sympathetic solar eruptions.

\acknowledgments SDO is a mission for NASA's Living With a Star (LWS) Program. The {\em STEREO}/SECCHI data used here were produced by an international consortium of the Naval Research Laboratory (USA), Lockheed Martin Solar and Astrophysics Lab (USA), NASA Goddard Space Flight Center (USA), Rutherford Appleton Laboratory (UK), University of Birmingham (UK), Max-Planck-Institut for Solar System Research (Germany), Centre Spatiale de Li$\rm \grave{e}$ge (Belgium), Institut d'Optique Th$\rm \acute{e}$orique et Appliqu$\rm \acute{e}$e (France), and Institut d'Astrophysique Spatiale (France). We thank the KSO, {\em RHESSI}, and {\em GOES} teams for data support. We would like to thank the anonymous referee for constructive comments and suggestions that improved the content of the manuscript. This work is supported by the Natural Science Foundation of China under grants 10933003, 11078004, and 11073050, and the National Key Research Science Foundation (2011CB811400), and Open Research Program of Key Laboratory of Solar Activity of National Astronomical Observatories of Chinese Academy of Sciences (KLSA2011\_14).


\begin{table}
\begin{center}
\caption{Time Line of the sub-events and their Corresponding Features\label{tb}}
\begin{tabular}{cl}
\tableline\tableline
Approx. Time & Corresponding feature \\
\tableline
11:30 UT &the first appearance of the plasma ejection\\
11:32 UT &the plasma ejection was split into two parts\\
11:40 UT &the southern part of the ejection acted on F2\\
11:45 UT &F2 started to rise slowly\\
11:56 UT &start of the small pre-flare hump, peaked at 12:00 UT\\
         &brightening appeared at around P1 and N1\\
         &brighten up of the loops that connect P1 and N1\\
         &an HXR source detected around the apex of the loops\\
12:03 UT &F1 was activated and started to rise\\
12:08 UT &the slow rising of F2 turned into the acceleration phase\\
         &the start of the main {\em GOES} C1.8 flare peaking at 12:28 and 12:37 UT\\
12:15 UT &the first appearance of the blob-like feature\\
12:33 UT &F1 reached the maximum erupting height and started to fall back\\
         &the start of the fast eruption of the blob-like feature\\
         &the acceleration of F2 transited into the fast eruption phase\\
\tableline
\end{tabular}
\end{center}
\end{table}

\end{document}